\documentclass[aps,twocolumn,prd,showpacs]{revtex4}
\usepackage[dvips]{graphicx}
\usepackage{graphics}
\usepackage{dcolumn}
\usepackage{amssymb}
\usepackage{bm}
\usepackage{amsmath}
\usepackage{color}

\def\spose#1{\hbox to 0pt{#1\hss}}

\def\lta{\mathrel{\spose{\lower 3pt\hbox{$\mathchar"218$}}
     \raise 2.0pt\hbox{$\mathchar"13C$}}}
\def\gta{\mathrel{\spose{\lower 3pt\hbox{$\mathchar"218$}}
     \raise 2.0pt\hbox{$\mathchar"13E$}}}

\newcommand{\dl}{\partial}
\newcommand{\fracc}[2]{\frac{\textstyle{#1}}{\textstyle{#2}}}
\newcommand{\be}{\begin{equation}}
\newcommand{\bea}{\begin{eqnarray}}
\newcommand{\ena}{\end{eqnarray}}

\def\setR{\mathbb{R}}

\def\ii{\textrm i}

\newcommand{\dbb}{de$\,$Broglie-Bohm}

\begin{document}
\title{Quantum-to-classical transition of primordial cosmological perturbations in \dbb\ quantum theory: the bouncing scenario}
\author{Nelson Pinto-Neto}
\affiliation{ICRA - Centro Brasileiro de
Pesquisas F\'{\i}sicas -- CBPF, rua Xavier Sigaud, 150, Urca,
CEP22290-180, Rio de Janeiro, Brazil}

\author{Grasiele Santos}
\affiliation{ICRA - Centro Brasileiro de
Pesquisas F\'{\i}sicas -- CBPF, rua Xavier Sigaud, 150, Urca,
CEP22290-180, Rio de Janeiro, Brazil}

\author{Ward Struyve}
\affiliation{Departments of Mathematics and Philosophy, Rutgers University, Hill Center, 110 Frelinghuysen Road, Piscataway, NJ 08854-8019, USA.}

\date{\today}

\begin{abstract}

In a previous work we have exhibited a clear description of the quantum-to-classical transition of cosmological quantum fluctuations in the inflationary scenario using the de Broglie-Bohm quantum theory. These fluctuations are believed to seed the small inhomogeneities, which are then responsible for the formation of large scale structures. In this work we show that using the \dbb\ theory it is also possible to describe the quantum-to-classical transition of primordial perturbations which takes place around a bouncing phase, even if the latter is caused by quantum effects due to the quantization of the background geometry.

\end{abstract}


\maketitle

\section{Introduction}
According to the usual accounts of structure formation, inflation drove the early Universe to a very homogeneous and isotropic configuration~\cite{liddle00,mukhanov05,weinberg08,lyth09,peter09}, in which only quantum vacuum fluctuations could have survived. These fluctuations have then resulted into classical perturbations of energy density which started to form structures (such as stars, galaxies and clusters of galaxies) through gravitational instability (although there are also models, known as warm inflation models, according to which those classical inhomogeneities originate from thermal fluctuations~\cite{berera95a,berera95b}). The same picture is advocated in some bouncing models~\cite{novello08}: small quantum fluctuations exist in a regime where the universe is dust-dominated, very big and rarefied. Their quantum state is very close to the Minkowski vacuum and the amplitudes of the fluctuations are amplified over time.

There has been a lot of discussion in the literature about the possible mechanisms behind the transition from cosmological quantum fluctuations to the small classical inhomogeneities which seed the structure formation~\cite{guth85,albrecht94,polarski96,lesgourgues97,kiefer98,kiefer09,burgess08,perez06,deunanue08,leon10,sudarsky11,martin12,canate13}. A successful explanation needs to deal with the measurement problem \cite{liddle00,mukhanov05,lyth09}. Namely, the initial quantum state is homogeneous and isotropic and needs to result in classical fluctuations which are not symmetric. The Schr\"odinger evolution preserves the translational and rotational symmetry. Hence, according to standard quantum theory, this symmetry can only be broken through collapse of the wave function. This collapse is supposed to happen upon measurement. But in the early universe there is no measurement device or observer which could cause such a collapse. Even worse, we are dealing with a description of the whole universe and hence 
there is not even place for an external measurement device or observer. All these structures, including measurement devices, are supposed to emerge from the primordial fluctuations themselves. As such, to address the problem of the quantum-to-classical transition of fluctuations, we need to use an alternative to quantum theory that is free of the measurement problem.

In a recent paper \cite{pinto-neto12}, we have considered the \dbb\ theory~\cite{bohm93,holland93b,durr09} to study the quantum-to-classical transition of these fluctuations for the case of inflationary theory (see \cite{hiley95} for an earlier study). We have shown that this transition can be explained very easily and naturally. The \dbb\ theory solves the measurement problem by postulating an actual field configuration, which is guided in its motion by the wave function. In the cosmological scenario, this actual field configuration is neither homogeneous nor isotropic, so it does not share the symmetries of the quantum state, and it behaves classically (i.e., obeys the classical field equations) when expected. In our analysis we did not appeal to decoherence. If there is suitable decoherence then our results still hold. In other approaches that solve the measurement problem, such as the many worlds theory, an explanation of the transition requires suitable decoherence. Much work has been performed on 
studying possible sources for the decoherence (it could come, for instance, from the coupling between different modes, or from interactions with other matter fields), as well as possible time scales at which it could occur, see e.g.\ \cite{burgess08} and references therein. However, it seems fair to say that there are no conclusive results yet concerning the source and time scales of the decoherence \cite{weinberg08}. 

Other approaches to the measurement problem worth mentioning are dynamical collapse theories. Various collapse models are being developed in order to account the quantum-to-classical transition of primordial fluctuations, see e.g.\ \cite{perez06,deunanue08,leon10,martin12,canate13} and references therein.

In this paper, we address the problem of the quantum-to-classical transition of cosmological perturbations for bouncing scenarios \cite{vitenti12}. In inflationary models, from the perspective of de Broglie-Bohm quantum theory, the fast suppression of the decaying mode and the consequent dominance of the growing mode ensured the quantum-to-classical transition of the perturbations. However, in bouncing models growing and decaying modes are interchanged at the bounce. Hence it is a priori not clear when the dominance of one mode over the other becomes really effective and it is not trivial to know when the perturbations will begin to behave classically.  Moreover, bounces can happen due to quantum gravitational effects~\cite{pinho07} and this could also disturb the classicality of the perturbations. We will show that, in spite of these facts, in general bouncing models, the perturbation modes begin to behave classically before the bounce takes place if the physical scale of the perturbations has become larger 
than the curvature scale of the background and the contraction has lasted a sufficiently long period of time. In fact, this is the case for all perturbations of cosmological interest. The calculations were done, again, in the context of the \dbb\ quantum theory.

In the next section we will summarize the classical dynamics of cosmological perturbations in general bouncing models. In section \ref{dbb}, we will consider the quantum description of the perturbations, in the context of the \dbb\ theory, and show how the quantum-to-classical transition takes place. In section \ref{quantumbounce} a particular example of a quantum bouncing model is presented. We end up with the conclusions in section \ref{conclusions}.

\section{\label{sec:pert}Linear cosmological perturbations in general bouncing models}
Let us first consider the classical description of cosmological perturbations. In the next section we will turn to the quantum description.

The perturbations are considered in a background Friedmann-Robertson-Walker model, with scale factor $a$ and uniform total matter distribution with density $\rho$ and pressure $p$, and are described by the Mukhanov-Sasaki variable $v({\bf x},\eta)$, which combines both fluctuations of the matter and metric. The parameter $\eta$ is conformal time defined by $a d\eta = dt$, $t$ being cosmic time. The Lagrangian for the Mukhanov-Sasaki variable, which can be derived from the Einstein-Hilbert action, is given by
\begin{equation}
\label{L}
L_v = \int d^3x\frac{1}{2}\left[v^{\prime2}+\left(\frac{z^\prime}{z}\right)^2v^2-c_s^2\delta^{ij}\dl_iv\dl_jv-2\frac{z^{\prime}}{z}vv^\prime\right],
\end{equation}
where the primes denote derivatives with respect to conformal time and
\begin{equation}\label{eq:zz}
z = \frac{\sqrt{\beta}}{x\mathcal{H}c_s}, \quad \beta = \frac{3}{2}\frac{8\pi \mathrm{G}}{3c^4}a^2\left(\rho+p\right), \quad c_s^2 = \frac{d p}{d\rho}.
\end{equation}
$\mathcal{H}= a^{\prime}/a $ is the conformal Hubble function, which relates to the Hubble function $H = a^{-1} da/ dt$ through $\mathcal{H} = a H$ and $x = a_0/a$ is the red-shift function. Subscripts 0 refer to present day values. The Lagrangian yields the following equations of motion for the Fourier modes $v_{\bf k}(\eta)$,
\begin{equation}
\label{eq:muk}
v^{\prime\prime}_{\bf k} + \left(c_s^2k^2-\frac{z^{\prime\prime}}{z}\right)v_{\bf k} = 0.
\end{equation}

Defining $\Omega = \rho/\rho_c$, where $\rho_c$ is the critical density today, and using the energy conservation equation
\begin{equation}
\label{conservation}
\frac{d \rho}{d t}+3H(\rho+p) = 0\;\;\Rightarrow\;\;
\frac{d\rho}{d x} = \frac{3\left(\rho + p\right)}{x},
\end{equation}
we obtain
\begin{eqnarray}
&\beta = \frac{1}{2x R^2_H}\frac{d \Omega}{d x},\qquad z^2 =  \frac{1}{2c_s^2x\Omega}\frac{d \Omega}{d x},\nonumber\\
&\quad c_s^2 = \frac{x}{3} \frac{d}{d x} \ln \left(\frac{1}{x^2}\frac{d \Omega}{d x}\right),
\end{eqnarray}
where $R_H =  c/(a_0 H_0)$ is the co-moving Hubble radius.

In the case the background matter is a single fluid with $p=w\rho$ we have $\Omega = \Omega_0 x^{3(1+w)}$ and
\begin{equation}\label{eq:cs2single}
c_s^2 = w, \quad  z^2 = \frac{3(1+w)}{2wx^2},
\end{equation}
and Eq.~\eqref{eq:muk} reduces to 
\begin{equation}
\label{eq:muk2}
v_{\bf k}^{\prime\prime} + \left(w k^2-\frac{a^{\prime\prime}}{a}\right)v_{\bf k} = 0.
\end{equation}

The general solution of the mode equation~\eqref{eq:muk} can be formally expanded in powers of
$k^2$ as~\cite{mukh92}
\begin{equation}
\begin{array}{l}
 \fracc{v_{\bf k}}{z} =  A_{1,{\bf k}}\biggl[1 - k^2 \displaystyle\int^{\eta}_{\eta_i} \fracc{d\bar
  \eta}{z^2\left(\bar \eta\right)} \displaystyle\int^{\bar{\eta}}
  c_s^2 z^2\left(\bar{\bar{\eta}}\right)d\bar{\bar{\eta}}+...\biggr]+
  \\ [2ex]A_{2,{\bf k}}\displaystyle\int^\eta_{\eta_i}\fracc{d\bar{\eta}}{z^2\left(\bar{\eta}\right)} \biggl[1 - k^2
  \displaystyle\int^{\bar{\eta}} c_s^2 z^2\left(\bar{\bar{\eta}}\right)
  d\bar{\bar{\eta}} \displaystyle\int^{\bar{\bar{\eta}}}
  \fracc{d\bar{\bar{\bar{\eta}}}}{z^2\left(\bar{\bar{\bar{\eta}}}\right)} + ...\biggr],\cr
\label{solform}
\end{array}
\end{equation}
where we have presented the terms up to order $ \mathcal{O}(k^{2})$. The lower bounds $\eta_i$ in the integrals are related to initial conditions that depend on the specific model being considered. The coefficients $A_{1,{\bf k}}$ and $A_{2,{\bf k}}$ are two constants, determined by the initial conditions, which are roughly the same order of magnitude~\footnote{In fact,
as $A_1$ and $A_2$ depend on ${\bf k}$, this assertion depends on the scale we are talking about. For an
almost scale invariant spectrum of cosmological perturbations, the $A_2$ term is larger than the $A_1$ term
for all scales of cosmological interest, see Ref.~\cite{pinho07},
which enforces the argumentation described below for the transition of quantum-to-classical behavior in
bouncing models. Only for very short wavelengths can the $A_1$ term be bigger than $A_2$.}. We will be interested
in the situation where $c_s^2 k^2\ll z''/z$ and hence we will take only the first terms of the series above.

This description of the perturbations is valid, in the case where entropy perturbations are negligible,
in the contracting and expanding phases when the dynamics is given by the General Relativity Einstein's equations,
and through the bounce itself in the case of the quantum bounce which we will discuss below.
For general bounces, it is not clear that solution (\ref{solform}) is valid through the bounce, neither do we have
any particular analytic solution in order to evaluate it away from the bounce as we have in the case of the quantum bounce presented in~\cite{pinho07}, which will be considered in section \ref{quantumbounce}. However, if the bounce is short enough, an estimate of Eq.~(\ref{solform}) away from the bounce, where General Relativity is valid, will be sufficient to evaluate the orders of magnitude of the amplitudes, assuming that a short bounce does not change the mode evolution to much.

In order to analyze the quantum-to-classical transition, we focus on the term
\begin{equation}
A_{2,{\bf k}}\int_{-\infty}^\eta\frac{d\bar\eta}{\bar z^2},
\end{equation}
which appears in the solution (\ref{solform}) (we are assuming that the contracting phase begins at a very large negative conformal time, which we take to be $\eta_i \rightarrow - \infty$). This term grows with time. We can write
\begin{equation}
\int_{-\infty}^\eta\frac{d\bar\eta}{z^2(\bar{\eta}) }=\left(B-\int_{\eta}^\infty\frac{d\bar \eta}{z^2(\bar \eta)}\right),
\end{equation}
where
\begin{equation}
B = \int_{-\infty}^\infty d\eta\,z^{-2}
\end{equation}
is a constant. For the case the bounce is dominated by a single fluid with equation of state parameter $w_q$, this constant was evaluated in Ref.~\cite{vitenti12} and reads
\begin{equation}\label{Bapprox}
B \approx \frac{4x_b}{3(1-w_q)E(x_b)z^2(x_b)},
\end{equation}
where $E = H/H_0=\sqrt{\Omega(x)}$, a subscript $b$ refers to the values of the physical quantities at the bounce.
It must be understood that, although evaluated at $x_b$, the functions $E(x)$ and
$z^2(x)$ in Eq.~(\ref{Bapprox}) are the usual general relativistic expressions for them
which are valid just before the bounce but maybe not through the bounce itself.

For realistic bounces occurring at energy scales bigger than the nucleosynthesis energy scale we have $x_b = a_0/a_b \gg 10^{10}$. Furtermore, since $w_q\ll1$, which is needed in order to obtain a scale invariant spectrum (see \cite{pinho07}), it follows that $B \gg 10^{10}$.
Hence, the solution for the mode functions $v_{\bf k}$ around the bounce ($\eta \gg - \infty$) is given by
\begin{eqnarray}
\label{modo-rico}
v_{\bf k} &\approx& [A_{1,{\bf k}} + A_{2,{\bf k}}B]z(\eta)-A_{2,{\bf k}}z(\eta)\int_\eta^\infty\frac{d\bar\eta}{z^2(\bar\eta)}\nonumber\\
&\approx&  A_{2,{\bf k}}z(\eta)\left[B - \int_\eta^\infty\frac{d\bar\eta}{z^2(\bar\eta)}\right].
\end{eqnarray}
In the last approximation we assumed that $A_1$ and $A_2$ are roughly of the same order. 

Equation (\ref{modo-rico}) is the main result of this section and will be used in the next one to achieve the classical limit. Remember that it is valid for perturbation modes for which $c^2_sk^2 \leqslant z''/z$ (i.e., when their physical wavelengths are much larger than the curvature scale of the contracting background) and in case the background space has already contracted enough.\\

\section{The \dbb\ approach to perturbations}\label{dbb}

We now consider a quantum mechanical treatment of the Mukhanov-Sasaki variable, keeping the background classical. The classical equation looks very similar to that of inflationary scenario. Therefore we can proceed in a similar way as in \cite{pinto-neto12}.

The Lagrangian (\ref{L}), which formally looks like that of a free scalar field with a time-dependent mass, can be straightfowardly quantized. Assuming a product wave functional $\Psi = \Pi_{{\bf k} \in \setR^{3+}} \Psi_{\bf k}(v_{\bf k},v^*_{\bf k},\eta)$, the Schr\"odinger equation for each wave function $\Psi_{\bf k}$ satisfies
\widetext
\begin{equation}
\label{sch}
\ii\frac{\partial\Psi_{\bf k}}{\partial\eta}=
\left[ -\frac{\partial^2}{\partial v_{\bf k}^*\partial v_{\bf k}}+
c^2_s k^2 v_{\bf k}^* v_{\bf k}
- \ii\frac{z'}{z}\left(\frac{\partial}{\partial v_{\bf k}^*}v_{\bf k}^*+
v_{\bf k}\frac{\partial}{\partial v_{\bf k}}\right)\right]\Psi_{\bf k}.
\end{equation}

In the \dbb\ approach, there is also an actual field $v(\eta,{\bf x})$ whose Fourier modes satisfy the guidance equations
\begin{equation}
\label{guidance}
v'_{\bf k}= \frac{\partial S_{\bf k}}{\partial v^*_{\bf k}}+\frac{z'}{z}v_{\bf k},  \quad
{v^*_{\bf k}}'= \frac{\partial S_{\bf k}}{\partial v_{\bf k}}+\frac{z'}{z}v^*_{\bf k} .
\end{equation}

Whenever $z''/z$ is negligible with respect to $c_s^2k^2$, which happens in the far past in the contracting phase for the modes of physical interest, vacuum initial conditions can be imposed on the wave function. This yields the solution
\begin{equation}
\label{psi2}
\Psi_{\bf k} = \frac{1}
{\sqrt{\sqrt{2\pi}|f_k(\eta)|}} \exp{\left\{-\frac{1}{2|f_k(\eta)|^2}|v_{\bf k}|^2 +   \ii \left[\left(\frac{|f_k(\eta)|'}{|f_k(\eta)|}-
\frac{z'}{z}\right)|v_{\bf k}|^2-
\int^\eta \frac{d {\tilde \eta}}{2|f_k({\tilde \eta})|^2}\right]\right\}} ,
\end{equation}
with $f_k$ a solution to the classical mode equation~\eqref{eq:muk} that is isotropic (so that $A_1$ and $A_2$ in \eqref{solform} depend only on $k$) and that satisfies $f_k(\eta_i) = 1/\sqrt{2k}$ with $|\eta_i|\gg1$ (so that $A_{1,{\bf k}} \equiv A_{1,k}= 1/\sqrt{2k}$). The quantum state $\Psi$ is homogeneous and isotropic. 

\endwidetext

For this quantum state, the solutions to the guidance equations read
\be
\label{soly}
v_{\bf k}(\eta) =  v_{\bf k}(\eta_i)\frac{|f_k(\eta)|}{|f_k(\eta_i)|}.
\end{equation}
Note that this result is independent of the precise form of $f_k(\eta)$. Since $f_k$ is a solution of the classical equation of motion (\ref{eq:muk}), the classical limit will be achieved whenever $|f_k(\eta)|$, as a function of time, is proportional to $f_k(\eta)$. As we have seen, around the bounce we have that (see Eq.~(\ref{modo-rico})),
\begin{equation}
f_k(\eta)\approx A_{2,k}z(\eta) \left[B - \int_\eta^\infty\frac{d\bar\eta}{z^2(\bar\eta)}\right],
\end{equation}
Thus, for large wavelengths, i.e., $c_s^2 k^2\ll z''/z$, we have $$f_k(\eta)\propto|f_k(\eta)|,$$
and hence
\begin{equation}
\label{new}
v_{\bf k}(\eta)\propto f_k(\eta),
\end{equation}
which means that the perturbations \eqref{soly} are evolving classically.

\section{Example of a quantum bounce}\label{quantumbounce}

As an example, we will present a quantum cosmological model with a perfect fluid with equation of state $p=\omega\rho$ modelling the matter source, quantized following the Wheeler-DeWitt prescription (see \cite{pinho07} for details). The background spatial metric is assumed to be flat. Considering only scalar perturbations, the Wheeler-DeWitt equation for $\Psi[a,v,T]$ reads
\begin{multline}
\label{universalwave}
i\frac{\partial \Psi}{\partial T}=\frac{1}{4}a^{(3\omega-1)/2}\frac{\partial}{\partial a}\left[a^{(3\omega-1)/2}\frac{\partial}{\partial a}\right]\Psi+ \\
-\frac{a^{3\omega-1}}{2}\int{d^3x\frac{\delta^2\Psi}{\delta v^2}}+\frac{a^{3\omega+1}\omega}{2}\int{d^3xv^{,i}v_{,i}\Psi},
\end{multline}
where the scale factor $a$ is the only remaining gravitational degree of freedom of the background, and $T$ is the fluid degree of freedom used to define time. The perturbations are described by the variable $v$, which is the gauge invariant Mukhanov-Sasaki variable.

We assume a product wave function of the form
\begin{equation}
\label{ansatz}
\Psi[a,v,T]=\Psi_{(0)}[a,T]\Psi_{(2)}[a,v,T],
\end{equation}
where $\Psi_{(0)}$ satisfies
\begin{equation}
\label{solpsi0}
\ii\frac{\partial \Psi_{(0)} }{\partial T}=\frac{a^{(3\omega-1)/2}}{4}\frac{\partial}{\partial a}\left[a^{(3\omega-1)/2}\frac{\partial}{\partial a}\right]\Psi_{(0)}.
\end{equation}
From the above equation we get the following continuity equation for $\rho(a,T)=a^{(1-3\omega)/2}|\Psi_{(0)}(a,T)|^2$:
\begin{equation}
\frac{\partial\rho}{\partial T}-\frac{\partial}{\partial a}\left[\frac{a^{3\omega-1}}{2}\frac{\partial S}{\partial a}\rho\right]=0,
\end{equation}
where $S$ is the phase of $\Psi_{(0)}$. Using the \dbb\ approach, we postulate an actual scale factor which obeys the guidance equation
\begin{equation}
\label{guidance-scale}
\frac{d a}{dT}=-\frac{a^{3\omega-1}}{2}\frac{\partial S}{\partial a},
\end{equation}
Using a gaussian with width $\sqrt{T_c/2}$ as the initial wave function (see Ref.~\cite{pinho07}), integration of this equation yields
\begin{equation}
\label{a_trajectory}
a(T)=a_b\left[1+\left(\frac{T}{T_c}\right)^2\right]^{\frac{1}{3(1-\omega)}},
\end{equation}
where $a_b$ is an integration constant. 

Changing the description to conformal time according to the relation $d\eta=[a(T)]^{3\omega-1}dT$, we get for $\Psi_{(2)}$ the equation
\begin{multline}
\label{schpsi2}
\ii\frac{\partial\widetilde \Psi_{(2)} }{\partial\eta}= \frac{1}{2} \int d^3x\Bigg(-\frac{\delta^2}{\delta v^2}+ \omega v_{,i}v^{,i} \\
- \ii\frac{a'}{a}\Big( \frac{\delta}{\delta v} v + v\frac{\delta}{\delta v}\Big)\Bigg) \widetilde \Psi_{(2)},
\end{multline}
where, as before, the primes denote derivative with respect to $\eta$ and $a(\eta)$ is the solution of the guidance equation (\ref{guidance-scale}), and ${\widetilde \Psi}_{(2)}(v,\eta)$ is related to the conditional wave function $\Psi_{(2)}(a(T),v,T)$.

In this case, the equation of motion for the variable $f_k(\eta)$ defined in Eq.\ (\ref{psi2}) (with $z$ replaced by $a$) is identical to Eq.\ (\ref{eq:muk2}), except for the fact that the quantity $a$ is now a given function of time, the Bohmian trajectory shown in (\ref{a_trajectory}). Therefore, solution (\ref{solform}), with $z$ replaced by $a$ and $c_s^2$ by $w$, is valid through the bounce itself and the constant $B$ is
given exactly by
\begin{equation}
 B=\frac{8 w}{9(1-w)^2 \sqrt{\Omega_{0}}}x_b^{3(1-w)/2}.
\label{B}
\end{equation}
Note that this exact calculation is in accordance with the estimate (\ref{Bapprox}). Hence $B$ is very large, Eqs.~(\ref{modo-rico}) and (\ref{new}) are valid, and the quantum-to-classical transition takes place.

\section{Conclusions}\label{conclusions}
Using the \dbb\ theory, we have established the quantum-to-classical transition of primordial perturbations for a large class of bouncing models. We also considered the particular example of a quantum bounce described in \cite{pinho07}, for which there is an analytical solution for perturbations during the bounce.

While the wave function of the perturbations is homogeneous and isotropic, the actual perturbation $v({\bf x},\eta)$ is not symmetric. It is a superposition of $A_{1,k}$ and $A_{2,k}$ modes. As we showed in section \ref{dbb}, during the bounce the $A_{1,k}$ mode becomes negligible compared to the $A_{2,k}$ mode (see Eq.~(\ref{modo-rico})), so that the mode $v_{\bf k}(\eta)$ and hence the field $v({\bf x},\eta)$ behave classically.

As in the inflationary case, we did not invoke decoherence. If at some stage in this transition there is decoherence in the field basis, this will not destroy the classical behavior of the fields.

\begin{acknowledgments}
G.B.\ Santos would like to thank Faperj of Brazil for financial support. N.\ Pinto-Neto would like to thank CNPq of Brazil for financial support. W.S.\ acknowledges support of a grant from the John Templeton Foundation. The opinions expressed in this publication are those of the authors and do not necessarily reflect the views of the John Templeton Foundation.
\end{acknowledgments}

\end{document}